\documentclass[10pt,letterpaper]{article}
\pdfoutput=1
\usepackage{opex3}
\usepackage{cite}
\usepackage{amsmath}
\usepackage{amssymb}
\usepackage{graphicx}
\usepackage[hang,loose]{subfigure}
\usepackage{enumerate}

\newcommand{\citeasnoun}[1]{Ref.~\citenum{#1}}

\newcommand{\ep}{\varepsilon}
\newcommand{\dep}{\Delta\varepsilon}
\renewcommand{\vec}[1]{\mathbf{#1}}

\newcommand{\figref}[1]{Fig.~\ref{fig:#1}}

\newcommand{\secref}[1]{Sec.~\ref{sec:#1}}

\newcommand{\defn}{\triangleq}
\renewcommand{\vec}[1]{\mathbf{#1}}

\newcommand{\eqnumref}[1]{(\ref{eq:#1})}
\renewcommand{\eqref}[1]{eq.~\eqnumref{#1}}
\newcommand{\Eqref}[1]{Equation~\eqnumref{#1}}
\newcommand{\eqreftwo}[2]{eqs.~\eqnumref{#1} and~\eqnumref{#2}}

\begin{document}

\title{Rigorous sufficient conditions\\ for index-guided modes\\ in microstructured dielectric waveguides}

\author{Karen K. Lee, Yehuda Avniel, and Steven G. Johnson}

\address{Research~Laboratory~of~Electronics, Massachusetts~Institute~of~Technology, 77~Massachusetts~Ave., Cambridge~MA~02139.}

\email{kylkaren@mit.edu} 



\begin{abstract}
We derive a sufficient condition for the existence of index-guided
modes in a very general class of dielectric waveguides, including
photonic-crystal fibers (arbitrary periodic claddings, such as ``holey
fibers''), anisotropic materials, and waveguides with periodicity
along the propagation direction.  This condition provides a rigorous
guarantee of cutoff-free index-guided modes in any such structure
where the core is formed by increasing the index of refraction
(e.g. removing a hole).  It also provides a weaker guarantee of
guidance in cases where the refractive index is increased ``on
average'' (precisely defined).  The proof is based on a simple
variational method, inspired by analogous proofs of localization for
two-dimensional attractive potentials in quantum mechanics.
\end{abstract}



\section{Introduction}

In this paper, we present rigorous sufficient conditions for the
existence of index-guided modes, including conditions for cutoff-free
modes, in a wide variety of dielectric waveguides---from ordinary
step-index fibers~\cite{Snyder83}, to photonic-crystal ``holey''
fibers~\cite{Russell03,Bjarklev03,Zolla05,Joannopoulos08}, and even
fiber-Bragg gratings~\cite{Ramaswami98} or other periodically
modulated waveguides~\cite{Elachi76,Fan95:periodicwvg,Joannopoulos08}.
The dispersion relations of such waveguides must almost always be
computed numerically, and so exact analytical theorems like the one
derived here provide a foundation of certainty that is not available
in any other way.  A rigorous theorem allows us to give a general
answer (although not a necessary condition) for questions such as: if
the waveguide core has a mixture of higher- and lower-index regions,
how much higher-index material is enough for cutoff-free guidance; and
under what conditions do photonic-crystal fibers, like step-index
fibers, have \emph{cutoff-free} guided modes?  The theorem provides an
absolute guarantee, with no calculation required, that strictly
increasing the refractive index to form the waveguide (e.g. filling in
a hole of a holey fiber) yields a cutoff-free guided mode.  It also
leads directly to necessary conditions for single-polarization fibers,
the subject of another manuscript currently in prepration.  Our work
extends an earlier proof of guided modes for homogeneous-cladding,
non-periodic, dielectric waveguides with isotropic~\cite{Bamberger90}
or anistropic~\cite{Urbach96} materials, and is closely related in
spirit to proofs of the existence of bound modes in two-dimensional
potentials for quantum mechanics~\cite{Yang89}.

\begin{figure}[t]
\begin{center}
\subfigure[Cross section of a waveguide (e.g. a conventional fiber) with a homogeneous cladding and an arbitrary-shape core.]{\includegraphics[width=0.273\columnwidth]{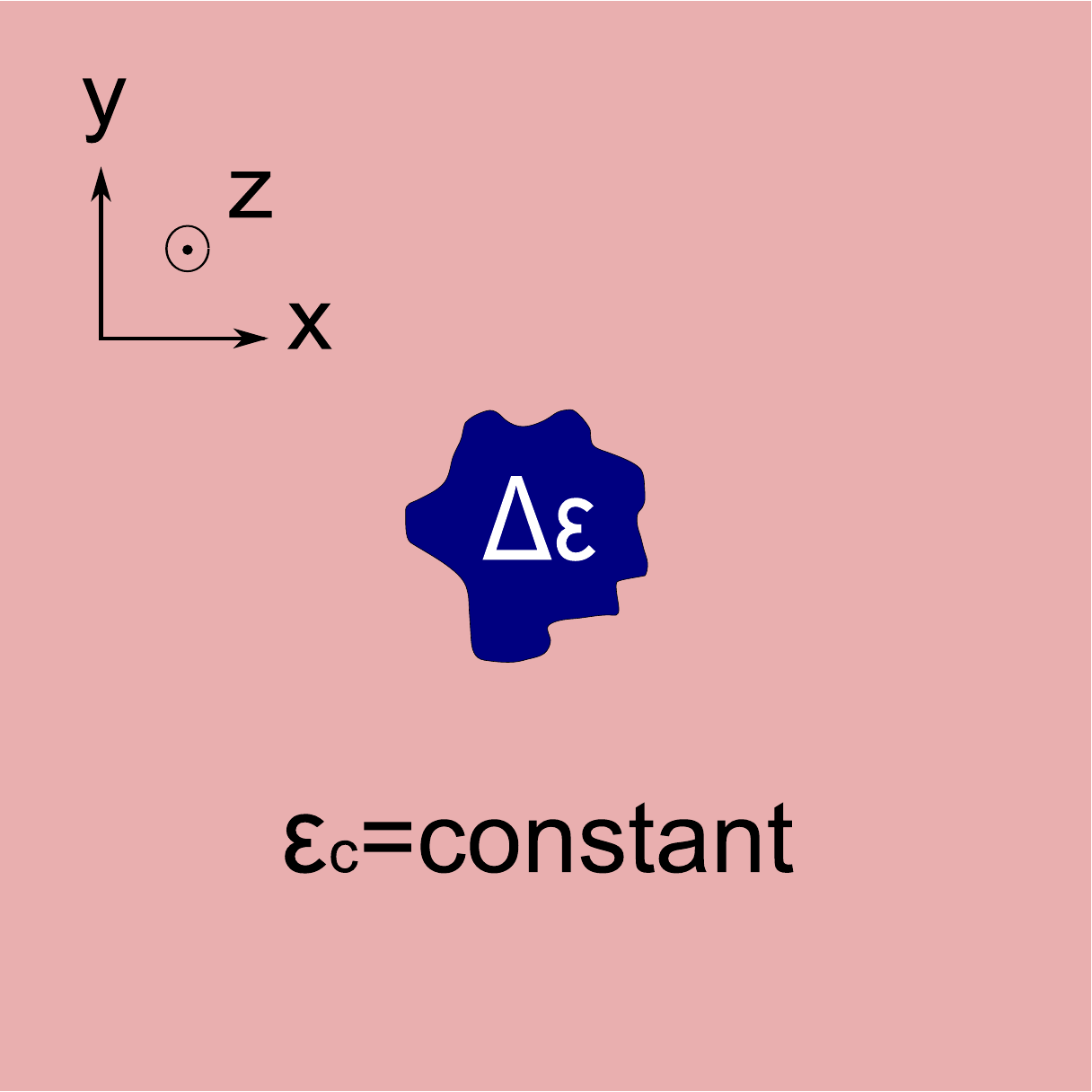}}
\subfigure[Cross section of a photonic-crystal fiber with periodic cladding and arbitrary-shape core.]{\includegraphics[width=0.3\columnwidth]{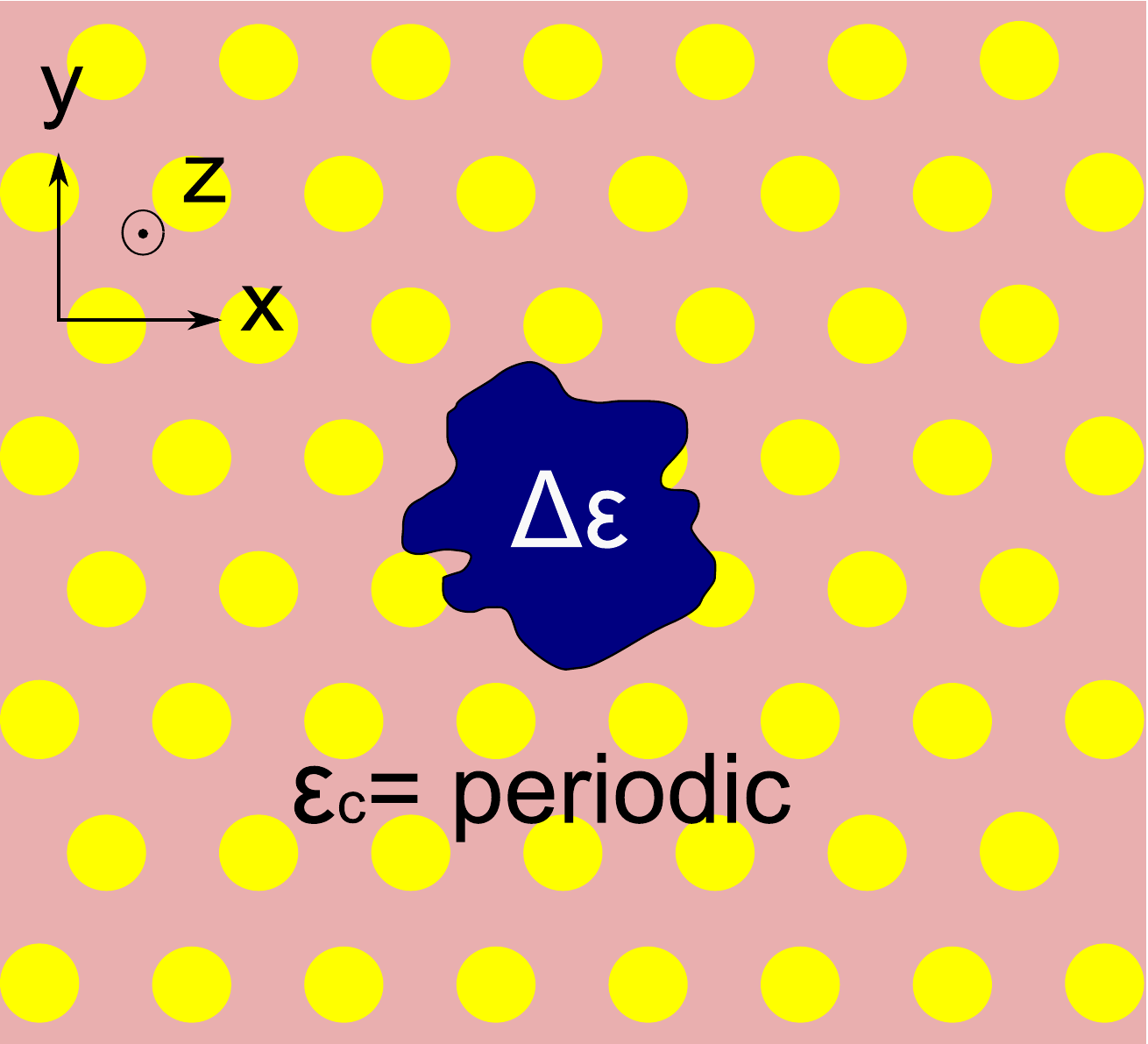}}
\subfigure[A waveguide periodic in the propagation ($z$) direction surrounded by a homogenous cladding.]{\includegraphics[width=0.27\columnwidth]{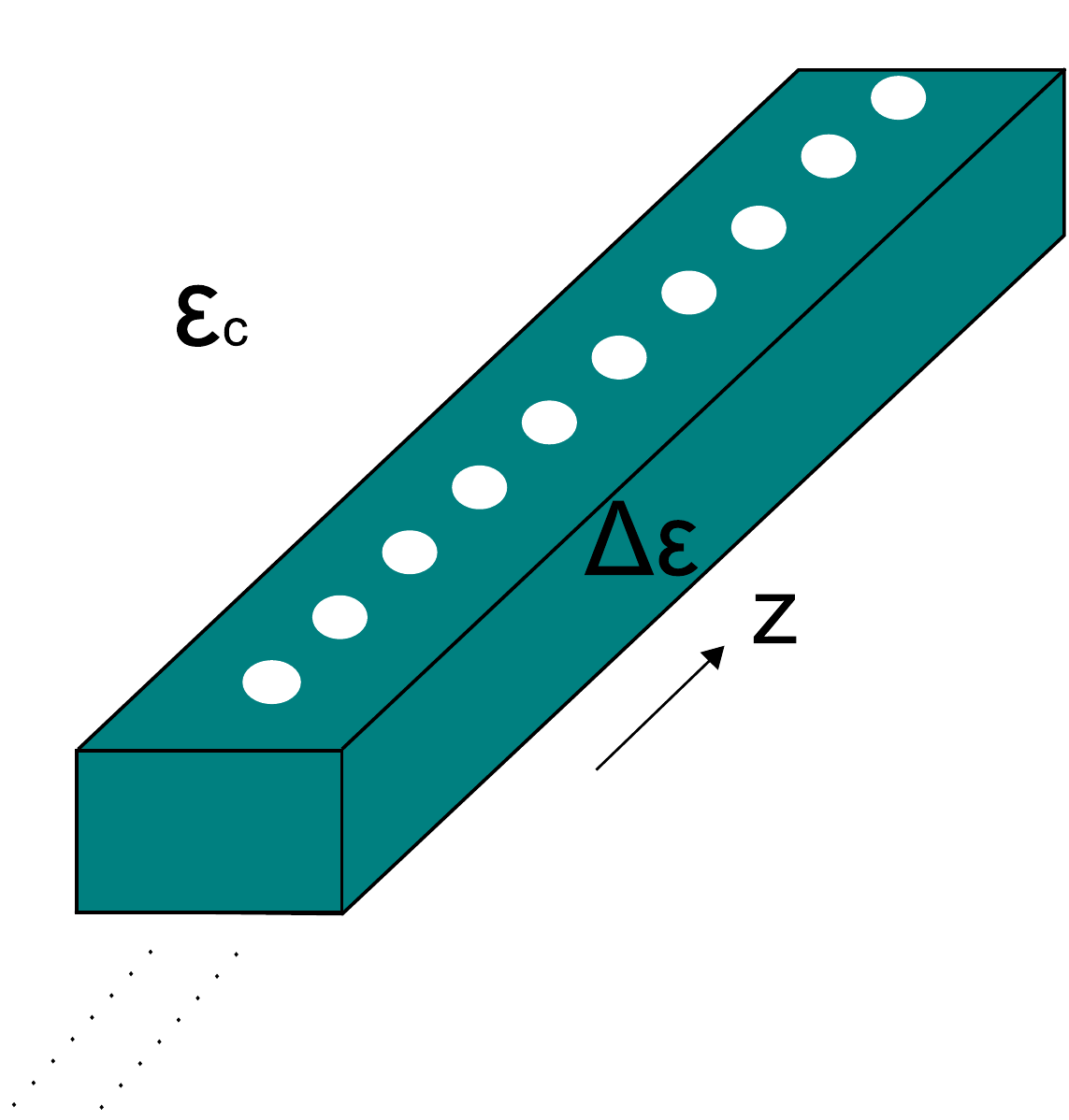}}
\caption{\label{fig:schematic}Schematics of various types of dielectric waveguides in which our theorem is applicable.  Light propagates in the $z$ direction (along which the structure is either uniform or periodic) and is confined in the $xy$ direction by a higher-index core compared to the surrounding (homogeneous or periodic cladding).}
\end{center}
\end{figure}

The most common guiding mechanism in dielectric waveguides is
\emph{index guiding} (or ``total internal reflection''), in which a
higher-index \emph{core} is surrounded by a lower-index
\emph{cladding} $\ep_c$ ($\ep$ is the relative permittivity, the
square of the refractive index in isotropic non-magnetic materials).
A schematic of several such dielectric waveguides is shown in
\figref{schematic}.  In particular, we suppose that the waveguide is
described by a dielectric function $\ep(x,y,z) = \ep_c(x,y,z) +
\dep(x,y,z)$ such that: $\ep$, $\ep_c$, and $\dep$ are periodic in $z$
(the propagation direction) with period $a$ ($a\to0$ for the common
case of a waveguide with a constant cross-section); that the cladding
dielectric function $\ep_c$ is periodic in $xy$ (e.g. in a
photonic-crystal fiber), with a homogeneous cladding (e.g. in a
conventional fiber) as a special case; and the core is formed by a
change $\dep$ in some region of the $xy$ plane, sufficiently localized
that $\int|1/\ep - 1/\ep_c|<\infty$ (integrated over the $xy$ plane
and the unit cell in $z$).  This includes a very wide variety of
dielectric waveguides, from conventional fibers
[\figref{schematic}(a)] to photonic-crystal ``holey'' fibers
[\figref{schematic}(b)] to waveguides with a periodic ``grating''
along the propagation direction [\figref{schematic}(c)] such as
fiber-Bragg gratings and other periodic waveguides.  We exclude
metallic structures (i.e, we require $\ep>0$) and make the
approximation of lossless materials (real $\ep$). We allow anisotropic
materials.  The case of substrates (e.g. for strip waveguides in
integrated optics~\cite{Hunsperger82, Saleh91, Chen06}) is considered
in \secref{substrates-etc}.  We also consider only non-magnetic
materials (relative permeability $\mu = 1$), although a future
extension to magnetic materials should be straightforward.
Intuitively, if the refractive index is increased in the core, i.e. if
$\dep$ is non-negative, then we might expect to obtain exponentially
localized index-guided modes, and this expectation is borne out by
innumerable numerical calculations, even in complicated geometries
like holey fibers~\cite{Russell03,Bjarklev03,Zolla05,Joannopoulos08}.

However, an intuitive expectation of a guided mode is far from a
rigorous guarantee, and upon closer inspection there arise a number of
questions whose answers seem harder to guess with certainty.  First,
even if $\dep$ is strictly non-negative, is there a guided mode at
\emph{every} wavelength, or is there the possibility of e.g. a
long-wavelength cutoff (as was initially suggested in holey
fibers~\cite{Kuhlmey02}, but was later contradicted by more careful
numerical calculations~\cite{Wilcox05})?  Second, what if $\dep$ is
\emph{not} strictly non-negative, i.e. the core consists of partly
increased and partly decreased index; it is known in such cases,
e.g. in ``W-profile fibers''~\cite{Kawakami74} that there is the
possibility of a long-wavelength cutoff for guidance, but precisely
how much decreased-index regions does one need to have such a cutoff?
Third, under some circumstances it is possible to obtain a
``single-polarization'' fiber, in which the waveguide is truly
single-mode (as opposed to two degenerate polarization modes as in a
cylindrical
fiber)~\cite{Okoshi80,Eickhoff82,Simpson83,Messerly91,Kubota04,Li05}---our
theorem can be extended, similar to \citeasnoun{Bamberger90}, to a
condition for \emph{two} guided modes, and we will explore the
consequences for single-polarization fibers in a subsequent paper.  It
turns out that all of these questions can be rigorously answered (in
the sense of sufficient conditions for guidance) for the very general
geometries considered in \figref{schematic}, without resorting to
approximations or numerical computations.

We will proceed as follows.  First, in \secref{theorem}, we review the
mechanism of index guiding, state our result (a sufficient condition
for the existence of index-guided modes), and discuss some important
special cases.  In \secref{homogeneous}, we first prove this theorem
for the simplified special case of a homogeneous cladding $\ep_c$,
where the proof is much easier to follow.  Then, in \secref{general},
we generalize the proof to arbitrary periodic claddings, such as for
holey photonic-crystal fibers (with some algebraic details left to the
appendix).  In \secref{substrates-etc}, we discuss a few contexts that
go beyond the initial assumptions of our theorem: substrates, material
dispersion, and finite-size effects.  Finally, we offer some
concluding remarks in \secref{conclusion} discussing future directions.

\section{Statement of the theorem}
\label{sec:theorem}

\begin{figure}[t]
\begin{center}
\includegraphics[width=0.7\columnwidth]{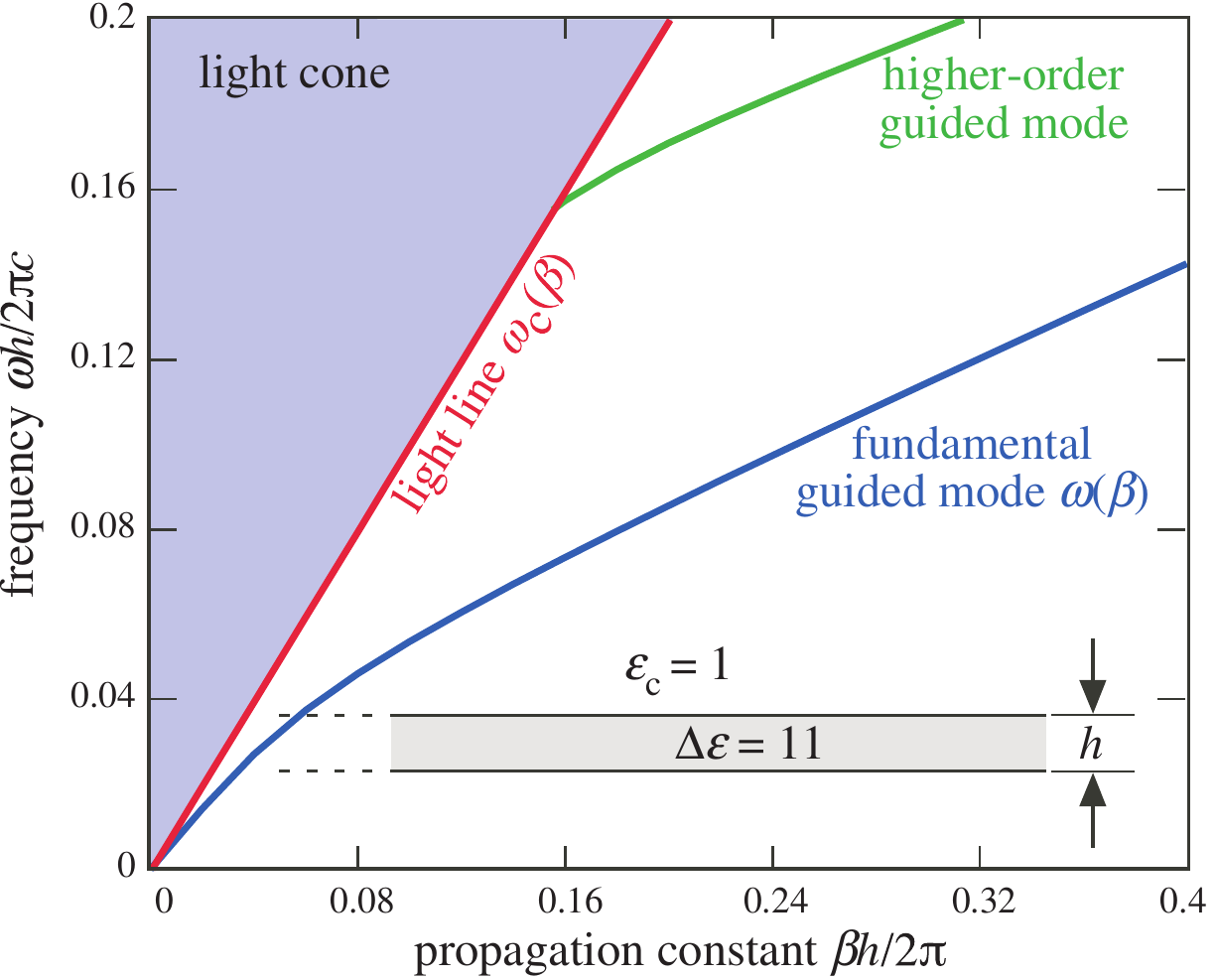}
\end{center}
\caption{\label{fig:banddiagram}Example dispersion relation of a
simple 2d dielectric waveguide in air (inset) for the TM polarization
(electric field out of the plane), showing the light cone, the light
line, the fundamental (cutoff-free) guided mode, and a higher-order
guided mode with a cutoff.}
\end{figure}

First, let us review the basic description of the eigenmodes of a
dielectric waveguide~\cite{Snyder83, Joannopoulos08}. In a waveguide
as defined above, the solutions of Maxwell's equations (both guided
and non-guided) can be written in the form of eigenmodes
$\vec{H}(x,y,z) e^{i\beta z - i\omega t}$ (via Bloch's theorem thanks
to the periodicity in $z$)~\cite{Joannopoulos08}, where $\omega$ is
the frequency, $\beta$ is the propagation constant, and the
magnetic-field envelope $\vec{H}(x,y,z)$ is periodic in $z$ with
period $a$ (or is independent of $z$ in the common case of a constant
cross section, $a\to0$).  A plot of $\omega$ versus $\beta$ for all
eigenmodes is the ``dispersion relation'' of the waveguide, one
example of which is shown in \figref{banddiagram}.  In the absence of the core
(i.e. if $\dep = 0$), the (non-localized) modes propagating in the
infinite cladding form the ``light cone'' of the
structure~\cite{Russell03,Bjarklev03,Zolla05,Joannopoulos08}; and at
each real $\beta$ there is a fundamental (minimum-$\omega$)
space-filling mode at a frequency $\omega_c(\beta)$ with a
corresponding field envelope
$\vec{H}_c$~\cite{Russell03,Bjarklev03,Zolla05,Joannopoulos08}.  Such
a light cone is shown as a shaded triangular region in \figref{banddiagram}.
Below the ``light line'' $\omega_c(\beta)$, the only solutions in the
cladding are evanescent modes that decay exponentially in the
transverse
directions~\cite{Kuchment01,Russell03,Bjarklev03,Zolla05,Joannopoulos08}.
Therefore, once the core is introduced ($\dep \neq 0$), any new
solutions with $\omega < \omega_c$ must be guided modes, since they
are exponentially decaying in the cladding far from the core: these
are the index-guided modes (if any).  Such guided modes are shown as
lines below the light cone in \figref{banddiagram}: in this case, both a
lowest-lying (``fundamental'') guided mode with no low-frequency
cutoff (although it approaches the light line asymptotically as
$\omega \to 0$) and a higher-order guided mode with a low-frequency
cutoff are visible.  Since a mode is guided if $\omega < \omega_c$, we
will prove the existence of a guided mode by showing that $\omega$ has
an upper bound $< \omega_c$, using the variational (minimax) theorem
for Hermitian eigenproblems~\cite{Joannopoulos08}.

[Modes that lie beneath the light light are not the only type of
guided modes in microstructured dielectric waveguides. While all the
guided modes in a traditional, homogeneous-cladding fiber lie below
the light line and are confined by the mechanism of index-guiding,
there can also be bandgap-guided modes in photonic-crystal
fibers~\cite{Russell03,Bjarklev03,Zolla05,Joannopoulos08}. These
bandgap-guided modes lie above the cladding light line and are
therefore not index-guided. Bandgap-guided modes always have a
low-frequency cutoff (since in the long-wavelength limit the structure
can be approximated by a ``homogenized'' effective medium that has no
gap~\cite{Smith05}).  We do not consider bandgap-guided modes in this
work; sufficient conditions for such modes to exist were considered
by~\citeasnoun{Kuchment04}.]

We will derive the following sufficient condition
for the existence of an index-guided mode in a dielectric waveguide at a
given $\beta$: a guided mode \emph{must} exist whenever
\begin{equation}
\int \vec{D}_c^* \cdot \left(\ep^{-1}-\ep_c^{-1}\right) \vec{D}_c < 0 ,
\label{eq:general-cond}
\end{equation}
where the integral is over $xy$ and one period in $z$ and $\vec{D}_c$
is the displacement field of the cladding's fundamental mode.  From
this, we can immediately obtain a number of useful special cases:
\begin{itemize}

\item There must be a cutoff-free guided mode if $\dep \geq 0$
everywhere (i.e., if we only increase the index to make the core).

\item For a homogeneous cladding (and isotropic media), there must be
a cutoff-free guided mode if $\int(1/\ep - 1/\ep_c) < 0$ (similar to
the earlier theorem of \citeasnoun{Bamberger90}, but generalized to
include waveguides periodic in $z$ and/or cores $\dep$ that do not have
compact support).

\item More generally, a guided mode has no long-wavelength cutoff if
\eqref{general-cond} is satisfied for the quasi-static
($\omega\to0$, $\beta\to0$) limit of $\vec{D}_c$.

\end{itemize}
\Eqref{general-cond} can also be extended to a sufficient condition
for having \emph{two} guided modes (or equivalently, a necessary
condition for single-polarization guidance), when the cladding
fundamental mode is doubly degenerate.  We explore this
generalization, analogous to a result in \citeasnoun{Bamberger90} for
homogeneous claddings, in another manuscript currently being prepared.

\section{Waveguides with a homogeneous cladding}  
\label{sec:homogeneous}

To illustrate the basic ideas of the proof in a simpler context, we
will first consider the case of a homogeneous cladding ($\ep_c =
\mathrm{constant}$) and isotropic, $z$-invariant strucures ($\ep$ is a
scalar function of $x$ and $y$ only).  In doing so, we reproduce a
result first proved (using a somewhat different approach) by
\cite{Bamberger90} (although the latter result required $\dep$ to have
compact support, whereas we only require a weaker integrability
condition).  Our proof, which we generalize in the next section, is
closely inspired by a proof~\cite{Yang89} of a related result in
quantum mechanics, the fact that any attractive potential in two
dimensions localizes a bound
state~\cite{Simon76,Landau77,Picq82,Yang89,Economou06}; we discuss
this analogy in more detail below.

That is, we take the dielectric function $\ep(x,y)$ to be of the form:
\begin{equation}
\ep(x,y) = \ep_c + \dep(x,y),
\end{equation}
where $\dep$ is an an arbitrary change in $\ep$ that forms the core of
the waveguide.  For convenience, we define a new function $\Delta$ by:
\begin{equation}
\Delta(x,y) \defn \ep^{-1} - \ep_c^{-1} .
\label{eq:Delta}
\end{equation}
The only constraints we place on $\dep$ are that $\ep$ be real and
positive and that $\int |\Delta| dxdy$ be finite, as discussed above.
Now, we wish to show that there must always be a (cutoff-free) guided
mode as long as $\dep$ is ``mostly positive,'' in the sense that:
\begin{equation}
\int \Delta(x,y) \, dxdy < 0,
\label{eq:homog-cond}
\end{equation}
Since \eqref{homog-cond} is independent of $\omega$ or $\beta$, the
existence of guided modes will hold at all frequencies
(cutoff-free).

The foundation for the proof is the existence of a variational
(minimax) theorem that gives an upper bound for the lowest
eigenfrequency $\omega_\mathrm{min}$.  In particular, at each
$\beta$, the eigenmodes $\vec{H}(x,y) e^{i\beta z - i\omega t}$
satisfy a Hermitian eigenproblem \cite{Joannopoulos08}:
\begin{equation}
\nabla_\beta \times \frac{1}{\ep} \nabla_\beta \times \vec{H} 
= \hat{\Theta}_\beta \vec{H} = \frac{\omega^2}{c^2} \vec{H},
\label{eq:eigen}
\end{equation}
where
\begin{equation}
\nabla_\beta \defn \nabla + i\beta\hat{\vec{z}} ,
\label{eq:nabla-beta}
\end{equation}
with \eqref{eigen} defining the linear operator
$\hat{\Theta}_\beta$. In addition to the eigenproblem, there is also
the ``transversality'' constraint~\cite{Kuchment01,Joannopoulos08}:
\begin{equation}
\nabla_\beta \cdot \vec{H} = 0
\label{eq:transversality}
\end{equation}
(the absence of static magnetic charges).  From the Hermitian property
of $\hat{\Theta}_\beta$, the variational theorem immediately
follows~\cite{Joannopoulos08}:
\begin{equation}
  \frac{\omega_\mathrm{min}^2(\beta)}{c^2} = \inf_{\nabla_\beta \cdot \vec{H} = 0}
  \frac{\int \vec{H}^* \cdot \hat{\Theta}_\beta \vec{H} dxdy}
       {\int \vec{H}^* \cdot \vec{H} dxdy}
\label{eq:variational-2d}
\end{equation}
That is, an upper bound for the smallest eigenvalue is obtained by plugging
\emph{any} ``trial function'' $\vec{H}(x,y)$, not necessarily an
eigenfunction, into the right-hand-side (the ``Rayleigh quotient''),
as long as $\vec{H}$ is ``transverse'' [satisfies
\eqref{transversality}].\footnote{Technically, we must also restrict
ourselves to trial functions where the integrals in
\eqref{variational-2d} are defined, i.e. the trial functions must
be in the appropriate Sobolev space $H(\nabla_\beta\times)$.}
Conversely, if \eqref{transversality} is not satisfied, it is easy to
make the numerator of the right-hand-side (which involved
$\nabla_\beta\times\vec{H}$) \emph{zero}, e.g. by setting
$\vec{H}=\nabla\varphi+i\beta\varphi\hat{\vec{z}}$ for any
$\varphi(x,y)$, so transversality of the trial function is critically
important to obtaining a true upper bound.

Now, we merely need to find a transverse trial function such that the
variational upper bound is below the light line of the cladding, which
will guarantee a guided fundamental mode.  For a homogeneous,
isotropic cladding $\ep_c$, the light line is simply $\omega_c^2/c^2 =
\beta^2 / \ep_c$, and so the condition for guided modes becomes:
\begin{multline}
\ep_c \int \vec{H}^* \cdot \hat{\Theta}_\beta \vec{H} dxdy
-  \beta^2 \int \vec{H}^* \cdot \vec{H} dxdy 
\\
= \ep_c \int \frac{1}{\ep} \left\| \nabla_\beta \times \vec{H} \right\|^2 dxdy
-  \beta^2 \int \left\| \vec{H} \right\|^2 dxdy 
< 0 ,
\label{eq:homog-var-condition}
\end{multline}
where in the second line we have integrated by parts.

The problem of bound states in quantum mechanics is conceptually very
similar.  There, given a potential function $V(x,y)$ in two dimensions
with $\int|V|<\infty$, one wishes to show that $\int V < 0$
(attractive) implies the existence of a bound state: an eigenfunction
of the Schr{\"o}dinger operator $-\nabla^2 + V$ with eigenvalue
(energy) $< 0$.  Again, this is a Hermitian eigenproblem and there is
a variational theorem~\cite{Tannoudji77}, so one merely needs to find
some trial wavefunction $\psi$ for which the Rayleigh quotient is
negative in order to obtain a bound state.  In one dimension, finding
such a trial function is simple---for example, an exponentially
decaying function $e^{-\alpha |x|}$ (or a Gaussian $e^{-\alpha x^2}$)
will work for sufficiently small $\alpha$---and the proof is sometimes
assigned as an undergraduate homework problem~\cite{Haar64}.  In two
dimensions, however, finding a trial function is more difficult---in
fact, no function of the form $f(\alpha r)$ (where $r$ is the radius
$\sqrt{x^2 + y^2}$) will work (without more knowledge of the explicit
solution for $V$)~\cite{Yang89}---and the earliest proofs of the existence of
bound modes used more complicated, non-variational
methods~\cite{Simon76, Economou06}.  However, an appropriate trial
function for a variational proof was eventually
discovered~\cite{Picq82,Bamberger90}, and later a simpler trial
function $e^{-(r+1)^\alpha}$ was proposed independently by Yang and
de~Llano~\cite{Yang89}.

\begin{figure}[t]
\begin{center}\includegraphics[width=0.7\columnwidth]{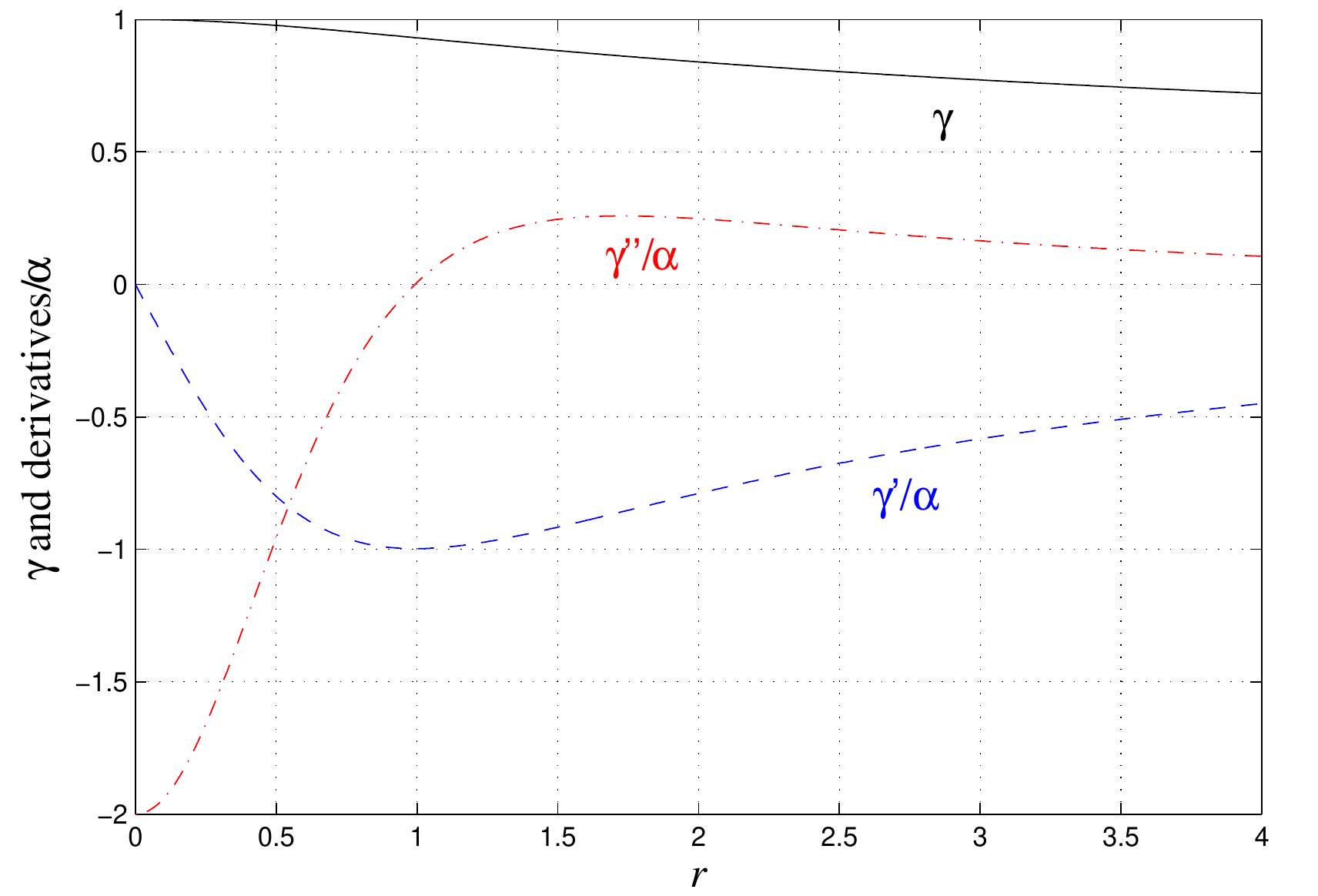}\end{center}
\caption{\label{fig:gamma}Plot of $\gamma$ [\eqref{gamma-def}],
$\gamma'/\alpha$ [\eqref{gammap}], and $\gamma''/\alpha$
[\eqref{gammapp}] versus $r$ for $\alpha=0.1$. All three functions go
to zero for $r\to\infty$, with no extrema other than those shown.}
\end{figure}

In the present electromagnetic case, we found that the following trial
function, inspired by the quantum case above~\cite{Yang89}, works.
That is, we can prove the existence of waveguided modes for a
homogeneous cladding using the trial function, in cylindrical
$(r,\phi)$ coordinates:
\begin{equation}
\vec{H} = \hat{\vec{r}} \gamma \cos\phi
          -\hat{\boldsymbol{\phi}} \left(r\gamma\right)' \sin\phi ,
\label{eq:homog-trial}
\end{equation}
where 
\begin{equation}
\gamma = \gamma(r) = e^{1-(r^2+1)^\alpha}
\label{eq:gamma-def}
\end{equation} 
for some $\alpha > 0$, and $(r\gamma)'$ is the derivative with respect
to $r$.  Clearly, $\vec{H}$ in \eqref{homog-trial} reduces to an
$\hat{\vec{x}}$-polarized plane wave propagating in the
$\hat{\vec{z}}$ direction as $\alpha\to0$ (and hence $\gamma\to1$).
This is a key property of the trial function: in the limit of no
localization ($\alpha = 0$, $\dep = 0$) it should recover a
fundamental (lowest-$\omega$) solution of the infinite cladding.
Also, by construction, it satisfies the transversality condition
\eqnumref{transversality} (which is why we chose this particular
form).  We chose $\gamma$ slightly differently from
\citeasnoun{Yang89} for convenience only (to make sure it is
differentiable at the origin and goes to $1$ for $\alpha\to 0$).  For
future reference, the first two $r$ derivatives of $\gamma$ are:
\begin{align}
\gamma' &= -2\alpha r(r^{2}+1)^{\alpha-1} \gamma, \label{eq:gammap} \\
\gamma'' &= 2\alpha (r^2 + 1)^{\alpha - 1} \gamma \left[-1 +
            2\alpha r^2(r^2+1)^{\alpha-1} + 2(1-\alpha)r^2(r^2+1)^{-1} 
             \right] , \label{eq:gammapp}
\end{align}
and are plotted along with $\gamma$ in \figref{gamma}.

What remains is, in principle, merely a matter of algebra to verify
that this trial function, for sufficiently small $\alpha$, satisfies
the variational condition \eqnumref{homog-var-condition}.  In
practice, some care is required in appropriately bounding each of the
integrals and in taking the limits in the proper order, and we 
review this process below.

We substitute the trial function \eqnumref{homog-trial} for $\vec{H}$
into the left-hand side of \eqref{homog-var-condition}:
\begin{equation}
\begin{split}
&\ep_c\int \frac{1}{\ep({\vec{r}})} \left\|\left(\nabla+i\beta\hat{z}\right)\times\vec{H}(r,\phi)\right\|^2 d^2\vec{r}-\beta^2\int \|\vec{H}\|^2 d^2\vec{r}\\
&=\ep_c\int\left(\frac{1}{\ep_c}+\Delta(\vec{r})\right)\left\|\hat{z} \frac{1}{r}\left[\frac{\partial}{\partial r}(rH_\phi)-\frac{\partial H_r}{\partial \phi}\right]+i\beta\hat{z}\times {\vec{H}}\right\|^2d^2\vec{r}-\beta^2\int {\|\vec{H}\|^2} d^2\vec{r}\\
&=\ep_c\int\left(\frac{1}{\ep_c}+\Delta(\vec{r})\right)\left(\frac{\sin^2\phi}{r^2}\left\{\left[r(r\gamma)'\right]'-\gamma\right\}^2+\beta^2\|\vec{H}\|^2\right)d^2\vec{r}-\beta^2\int {\|\vec{H}\|^2} d^2\vec{r}\\
&= \ep_c\int \left(\frac{1}{\ep_c}+\Delta(\vec{r})\right)
             \frac{\sin^2\phi}{r^2}\left(3r\gamma'+r^2\gamma''\right)^2 d^2\vec{r}
 + \ep_c\int\beta^2\Delta(\vec{r}) \|\vec{H}\|^2d^2\vec{r}\\
\end{split}
\label{eq:homog-var-condition2}
\end{equation}
We proceed to show that the last line of the above expression is negative in
the limit $\alpha\to0$, thus satisfying the condition for the
existence of bound modes.  We first examine the second term of
\eqref{homog-var-condition2}:
\begin{equation}
\lim_{\alpha\to{0}}\int\beta^2\Delta(\vec{r})\|\vec{H}\|^2 d^2\vec{r} =
\beta^2 \int \Delta(\vec{r}) d^2\vec{r} < 0 .
\label{eq:uniform-conv}
\end{equation}
The key fact here is that we are able to interchange the $\alpha\to0$
limit and the integral in this case, thanks to Lebesgue's Dominated
Convergence Theorem (LDCT)~\cite{Hewitt65}: whenever the absolute value
of the integrand is bounded above (for sufficiently small $\alpha$) by
an $\alpha$-independent function with a finite integral, LDCT guarantees
that the $\alpha\to0$ limit can be interchanged with the integral.  In
particular, the absolute value of this integrand is bounded above by
$|\Delta|$ multiplied by some constant (since $|\vec{H}|$ is bounded by
a constant: $|\gamma|\le 1$ and $|r\gamma'|$ is also easily seen to be
bounded above for sufficiently small $\alpha$), and $|\Delta|$ has a
finite integral by assumption.  Since $\lim_{\alpha\to0} |\vec{H}|^2 =
1$, we obtain \eqref{homog-cond}, which is negative by assumption.

Now we must show that the remaining first term of
\eqref{homog-var-condition2} goes to zero as $\alpha\to0$, completing
our proof.  This term is proportional to $(\ep_c^{-1} + \Delta)$, but
the $\Delta$ terms trivially go to zero by the same arguments as
above: $\Delta$ allows the limit to be interchanged with the
integration by LDCT, and as $\alpha\to0$ the $\gamma'$ and $\gamma''$ terms go
to zero.  The remaining $\ep_c^{-1}$ terms can be bounded above by a
sequence of inequalities as follows:
\begin{equation}
\begin{split}
&\lim_{\alpha\to 0}\int_0^\infty \int_0^{2\pi}
\frac{\sin^2\phi}{r^2}(3r\gamma'+r^2\gamma'')^2 r\,dr\,d\phi
\\
&= 16\pi
\lim_{\alpha\to 0} \int_0^\infty \alpha^2r^3(r^2+1)^{2\alpha-2}\gamma^2\left[-2+\alpha r^2(r^2+1)^{\alpha-1}
+(1-\alpha)r^2(r^2+1)^{-1}\right]^2dr
\\
&\le 16\pi
\lim_{\alpha\to 0} \int_0^\infty \alpha^2 r (r^2+1)^{2\alpha-1}\gamma^2\left[2+\alpha (r^2+1)^{\alpha}
+(1-\alpha)\right]^2dr
\\
&= 16\pi\lim_{\alpha\to 0} \int_1^\infty \alpha^2 t^{4\alpha-1}
      e^{2-2t^{2\alpha}} \left[ (3-\alpha) + \alpha t^{2\alpha} \right]^2 dt
\\
&\le 8\pi \lim_{\alpha\to 0}\int_0^\infty \alpha u \, e^{2-2u} 
                        \left[(3-\alpha) + \alpha u\right]^2 du
\\
&= 8\pi e^2\lim_{\alpha\to 0} \alpha \left[\frac{3}{8}\alpha^2
+\frac{1}{2}\alpha(3-\alpha)+\frac{1}{4}\left(3-\alpha\right)^2\right] = 0 .
\end{split}
\label{eq:gamma-integral-limit}
\end{equation}
From the first to second line, we substituted
\eqreftwo{gammap}{gammapp} and simplified.  From the second to third
line, we bounded the integrand above by flipping negative terms into
positive ones and replacing $r^2$ with $r^2+1$.  From the third to the
fourth line, we made a change of variables $t^2=r^2+1$. Then, from the
fourth to fifth line, we made another change of variable
$u=t^{2\alpha}$, and bounded the integral above by changing the lower
limit from $u=1$ to $u=0$. The final integral can be performed exactly
and goes to zero, completing the proof.

\section{General periodic claddings}
\label{sec:general}

In the previous section we considered $z$-invariant waveguides with a
homogeneous cladding and isotropic materials (for example,
conventional optical fibers). We now generalize the proof in three ways,
by allowing:
\begin{itemize}
\item transverse periodicity in the cladding material (photonic-crystal fibers),
\item a core and cladding that are periodic in $z$ with period $a$
($a\to0$ for the $z$-invariant case),
\item anisotropic $\ep_c$ and $\dep$ materials ($\ep$ is a $3\times3$
positive-definite Hermitian matrix).
\end{itemize} 
In particular, we consider dielectric functions of the form:
\begin{equation}
\ep(x,y,z) = \ep_c(x,y,z) + \dep(x,y,z),
\end{equation}
where the cladding dielectric tensor $\ep_c(x,y,z)=\ep_c(x,y,z+a)$ is
$z$-periodic and also periodic in the $xy$ plane (with an arbitrary
unit cell and lattice), and the core dielectric tensor change
$\dep(x,y,z)=\dep(x,y,z+a)$ is $z$-periodic with the \emph{same}
period $a$. Both $\ep_c$ and the total $\ep$ must be
positive-definite Hermitian tensors. As defined in \eqref{Delta}, we
denote by $\Delta$ the change in the inverse dielectric tensor.
Similar to the isotropic case, we require that $\int|\Delta_{ij}|$ be
finite for integration over the $xy$ plane and one period of $z$, for
every tensor component $\Delta_{ij}$.  We also require that
the components of $\ep_c^{-1}$ be bounded above.

In the homogeneous-cladding case, any light mode that lies beneath the
(linear) light line of the cladding is guided. We have shown that such
a mode \emph{always} exists, for all $\beta$, under the condition of
\eqref{homog-cond}, by showing that the variational upper bound on its
frequency lies below the light line. In the case of a periodic
cladding, the light line is the dispersion relation of the fundamental
space-filling mode of the cladding, which corresponds to the
lowest-frequency mode at each given propagation constant $\beta$
\cite{Russell03,Bjarklev03,Zolla05,Joannopoulos08}. This light ``line'' is,
in general, no longer straight, and there are mechanisms for guidance
that are not available in the previous case, such as bandgap
guidance~\cite{Russell03,Bjarklev03,Zolla05,Joannopoulos08}. Bandgap-guided
modes may exist above the light line and are, in general, not
cutoff-free because the gap has a finite bandwidth. Here, we
\emph{only} consider index-guided modes, which are guided because they
lie below the light line. We will follow the same general procedure as
in the previous section to derive the sufficient condition
[\eqref{general-cond}] to guarantee the existence of guided modes.
The homogeneous-cladding case is then a special case of this more
general theorem, recovering \eqref{homog-cond} (but generalizing it to
$z$-periodic cores), where in that case the cladding fundamental mode
$\vec{D}_c$ is a constant and can be pulled out of the integral.  The
case of a $z$-homogeneous fiber is just the special case $a\to0$,
eliminating the $z$ integral in \eqref{general-cond}.

The proof is similar in spirit to that of the homogeneous-cladding
case. At each $\beta$, the eigenmodes $\vec{H}(x,y,z) e^{i\beta z
-i\omega t}$ satisfy the same Hermitian eigenproblem \eqnumref{eigen}
and transversality constraint \eqnumref{transversality} as before. We
have a similar variational theorem to
\eqref{variational-2d}~\cite{Joannopoulos08}, except that, in the case
of $z$-periodicity, we now integrate over one period in $z$ as well as
over $x$ and $y$.
\begin{equation}
\label{eq:variational}
  \frac{\omega_\mathrm{min}^2(\beta)}{c^2} = \inf_{\nabla_\beta \cdot \vec{H} = 0}
  \frac{\int\vec{H}^* \cdot \hat{\Theta}_\beta \vec{H}}
       {\int\vec{H}^* \cdot \vec{H}}.
\end{equation}
As before, to prove the existence of a guided mode we will find a
trial function $\vec{H}$ such that this upper bound, called the
``Rayleigh quotient'' for $\vec{H}$, is below the light line
$\omega_c(\beta)^2 / c^2$.  The corresponding condition on $\vec{H}$
can be written [similar to \eqref{homog-var-condition}]:
\begin{equation}
\int \vec{H}^* \cdot \hat{\Theta}_\beta \vec{H}
-  \frac{\omega_c^2(\beta)}{c^2} \int \vec{H}^* \cdot \vec{H} < 0.
\label{eq:gen-var-condition}
\end{equation}

We considered a variety of trial functions, inspired by the Yang and
de~Llano quantum case~\cite{Yang89}, before finding the following
choice that allows us to prove the condition
\eqnumref{gen-var-condition}.  Similar to \eqref{homog-trial}, we want
a slowly decaying function proportional to $\gamma(r) =
e^{1-(r^2+1)^\alpha}$, from \eqref{gamma-def}, that in the
$\alpha\to0$ (weak guidance) limit approaches the cladding fundamental
mode $\vec{H}_c$.  As before, the trial function must be transverse
($\nabla_\beta \cdot \vec{H} = 0$), which motivated us to write the
trial function in terms of the corresponding vector potential. We
denote by $\vec{A}_c$ the vector potential corresponding to the
cladding fundamental mode $\vec{H}_c = \nabla_\beta \times \vec{A}_c$.
In terms of $\vec{A}_c$ and $\gamma$, our trial function is then:
\begin{equation}\label{eq:general-trial}
\vec{H}=\nabla_\beta\times
\left(\gamma\vec{A}_c\right)=\gamma\vec{H}_c+\nabla\gamma\times
\vec{A}_c .
\end{equation}
For convenience, we choose $\vec{A}_c$ to be Bloch-periodic (like
$\vec{H}_c$, since $\vec{A}_c$ also satisifies a periodic Hermitian
generalized eigenproblem and hence Bloch's theorem
applies).\footnote{Alternatively, it is straightforward to show that
the Coulomb gauge choice, $\nabla_\beta \cdot \vec{A}_c = 0$, gives a
Bloch-periodic $\vec{A}_c$, by explicitly constructing the
Fourier-series components of $\vec{A}_c$ in terms of those of
$\vec{H}_c$.} In contrast, our previous homogeneous-cladding trial
function [\eqref{homog-trial}] corresponds to a different gauge choice
with an unbounded vector potential $\vec{A}_c = -\frac{1}{i\beta}\hat{\vec{y}} +
\nabla_\beta\psi$, differing from a constant vector potential via the
gauge function $\psi = \frac{r}{i\beta}\sin\phi+e^{-i\beta z}$.

Substituting \eqref{general-trial} into the left hand side of our new
guidance condition \eqnumref{gen-var-condition}, we obtain five
categories of terms to analyze:
\begin{enumerate}[(i)]
\item terms that contain $\Delta=\ep^{-1}-\ep_c^{-1}$,
\item terms that cancel due to the eigenequation \eqnumref{eigen},
\item terms that have one first derivative of $\gamma$,
\item terms that have $(\gamma')^2$,
\item terms that have $\gamma'\gamma''$ or $(\gamma'')^2$.
\end{enumerate}
Category~(i) will give us our condition for guided modes,
\eqref{general-cond}, while category~(ii) will be cancelled exactly in
\eqref{gen-var-condition}.  We show in the appendix that all of the
terms in category~(iii) exactly cancel one another.  The terms in
categories (iv) and (v) all vanish in the $\alpha\to0$ limit; we
distinguish them because category~(v) turns out to be easier to
analyze.  There are no terms with $\gamma''$ alone, as these can be
integrated by parts to obtain category (iii) and (iv) terms.  In the
appendix, we provide an exhaustive listing of all the terms and how
they combine as described above.  In this section, we only outline the
basic structure of this algebraic process, and explain why the
category (iv) and (v) terms vanish as $\alpha\to0$.

Category~(i) consists only of one term:
\begin{equation}
\begin{split} 
&\lim_{\alpha\rightarrow 0}\int \vec{H}^*\cdot\left(\nabla_{\beta}\times \Delta \nabla_{\beta}\times \vec{H}\right)\\
=&\int \vec{H}_c^*\cdot\left(\nabla_{\beta}\times \Delta \nabla_{\beta}\times \vec{H}_c\right)\\
=&\int \left(\nabla_{\beta}\times \vec{H}_c\right)^*\cdot\Delta \left(\nabla_{\beta}\times \vec{H}_c\right)\\
=& \frac{\omega_c^2}{c^2}\int \vec{D}_c^*\cdot\Delta \vec{D}_c\\
\end{split}
\label{eq:uniform-conf-general}
\end{equation}
From the first to the second line, we interchanged the limit with the
integration, thanks to the LDCT condition as in \secref{homogeneous},
since the magnitudes of all of the terms in the integrand are bounded
above by the tensor components $|\Delta_{ij}|$ multiplied by some
$\alpha$-independent constants, and $|\Delta_{ij}|$ has a finite
integral by assumption.  (In particular, the $\vec{A}_c$ fundamental
mode and its curls are bounded functions, being Bloch-periodic, and
$\gamma$ and its first two derivatives are bounded for sufficiently
small $\alpha$.)  The result is precisely the left-hand side of
\eqref{general-cond}, which is negative by assumption.

Next, we would like to cancel $-\frac{\omega_c^2}{c^2}\int
\vec{H}^*\cdot \vec{H}$ by the eigen-equation \eqnumref{eigen}. Thus,
we examine the term $\int
\vec{H}^*\cdot\left(\nabla_{\beta}\times\ep_c^{-1}\gamma\nabla_{\beta}\times{\vec{H}_c}\right)$
(which comes from the term where the right-most curl falls on
$\vec{H}_c$ rather than $\gamma$) below:
\begin{equation}
\begin{split}
&\int \vec{H}^*\cdot \left(\nabla_{\beta}\times \ep_c^{-1}\gamma\nabla_{\beta}\times{\vec{H}_c}\right) \\
=&\int \vec{H}^*\cdot \left(\gamma\nabla_{\beta}\times\ep_c^{-1}\nabla_{\beta}\times \vec{H}_c+\left(\nabla\gamma\right)\times \ep_c^{-1}\nabla_{\beta}\times \vec{H}_c\right)\\
=&\int \vec{H}^*\cdot \gamma\frac{\omega_c^2}{c^2} \vec{H}_c+\int \vec{H}^*\cdot\left(\nabla\gamma\times \ep_c^{-1}\nabla_{\beta}\times \vec{H}_c\right)\\
=&\int \vec{H}^*\cdot \frac{\omega_c^2}{c^2} \vec{H}-\int \vec{H}^*\cdot \frac{\omega_c^2}{c^2}\nabla\gamma\times \vec{A}_c+\int \vec{H}^*\cdot\left(\nabla\gamma\times \ep_c^{-1}\nabla_{\beta}\times \vec{H}_c\right)\\
\end{split}
\label{eq:eigen-cancellation}
\end{equation}
From the second to the third lines, we used the eigenequation
\eqnumref{eigen}, and from the third to the fourth lines we used the
definition \eqnumref{general-trial} of $\vec{H}$ in terms of
$\vec{H}_c$.  The first term of the last line above cancels
$-\frac{\omega_c^2}{c^2}\int \vec{H}^*\cdot \vec{H}$ in
\eqref{gen-var-condition}. The second and third terms contain two
category~(iii) terms: $\frac{\omega_c^2}{c^2}\int \gamma
\vec{H}_c\cdot\left(\nabla\gamma\times\vec{A}_c\right)$ and
$i\omega_c \int \gamma\nabla\gamma \cdot \vec{E}_c \times
\vec{H}_c^*$, both of which will be exactly cancelled as described in
the appendix, as well as some category (iv) and (v) terms.

The category~(iv) integrands are all of the form $(\gamma')^2$
multiplied by some bounded function (a product of the various
Bloch-periodic fields as well as the bounded $\ep_c^{-1}$).  This
integrand can then be bounded above by replacing the bounded function
with the supremum $B$ of its magnitude, at which point the integral is
bounded above by $2\pi B \int_0^\infty (\gamma')^2 r\,dr$.  However,
such integrands were among the terms we already analyzed in the
homogeneous-cladding case, in \eqref{gamma-integral-limit}, and we
explicitly showed that such integrals go to zero as $\alpha\to0$.

The category~(v) integrands could also be explicitly shown to vanish
as $\alpha\to0$, similar to \eqref{gamma-integral-limit}, but a
simpler proof of the same fact can be constructed via the LDCT
condition.  In particular, similar to the previous paragraph, after
replacing bounded functions with their suprema we are left with
cylindrical-coordinate integrands of the form $\gamma'\gamma''r$ and
$(\gamma'')^2 r$.  Both of these integrands, however, are bounded
above by an $\alpha$-independent function with a finite integral, and
hence LDCT allows us to put the $\alpha\to0$ limit inside the integral
and set the integrands to zero.  Specifically, by inspection of
\eqreftwo{gammap}{gammapp}, $|\gamma'\gamma''|r < 4r^2(1+2+2) /
(r^2+1)^{2-\delta}$ and $(\gamma'')^2 r < 4r (1+2+2)^2 /
(r^2+1)^{2-\delta}$ for $\alpha<\delta/4$, and both of these upper
bounds have finite integrals, if we take $\delta$ to be some number $<
1/2$, since they decay faster than $1/r$.

In summary, we have shown that, if \eqref{general-cond} is satisfied,
then the variational upper bound for our trial function
[\eqref{general-trial}] is below the light line, and therefore an
index-guided mode is guaranteed to exist.  The special cases of this
theorem, as discussed in the introduction, immediately follow.  

\section{Substrates, dispersive materials, and finite-size effects}
\label{sec:substrates-etc}

In this section, we briefly discuss several situations that lie
outside of the underlying assumptions of our theorem: waveguides
sitting on substrates, dispersive ($\omega$-dependent) materials, and
finite-size claddings.

An optical fiber is completely surrounded by a single cladding
material, but the situation is quite different in integrated optical
waveguides.  There, it is common to have an asymmetrical cladding,
with air above the waveguide and a low-index material (e.g. oxide)
below the waveguide, such as in strip or ridge waveguides~\cite{Hunsperger82, Saleh91, Chen06}.
In such cases, it is well known that the fundamental guided mode has a
low-frequency cutoff even when the waveguide consists of strictly
nonnegative $\dep$~\cite{Hunsperger82, Chen06}.  This does not contradict our theorem
because we required the cladding to be periodic in both transverse
directions, whereas a substrate is not periodic in the vertical
direction.

We have also assumed non-dispersive materials in our proof. What
happens when we have more realistic, dispersive materials? Suppose
that $\ep$ depends on $\omega$ but has negligible absorption (so that
guided modes are still well-defined). For a given $\omega$, we can
construct a frequency-independent $\ep$ structure matching the actual
$\ep$ at that $\omega$, and apply our theorem to determine whether
there are guided modes at $\omega$.  The simplest case is when
$\dep\geq0$ for all $\omega$, in which case we must still obtain
cutoff-free guided modes.  The theorem becomes more subtle to apply
when $\dep < 0$ in some regions, because not only must one perform the
integral of \eqref{general-cond} to determine the existence of guided
modes, but the condition~\eqnumref{general-cond} is for a fixed
$\beta$ while the integrand is for a given frequency, and the
frequency of the guided mode is unknown {\it a priori}.

Finally, any real structure has a finite cladding.  Both numerically
and experimentally, this makes it difficult to study the
long-wavelength regime because the modal diameter increases rapidly
with wavelength (i.e. the frequency approaches the light line and the
transverse decay rate becomes very slow)---in fact, it seems likely
that the modal diameter will increase \emph{exponentially} with the
wavelength.  In quantum mechanics (scalar waves) with a potential well
of depth $V$, the decay length of the bound mode increases as
$e^{C/V}$ when $V\to0$, for some constant $C$~\cite{Simon76,Yang89}.
In electromagnetism, for the long wavelength limit, a homogenized
effective-medium $\tilde{\ep}$ description of the structure becomes
applicable~\cite{Smith05}, and in this effective near-homogeneous
limit the modes are described by a scalar wave equation with a
``potential'' $-\omega^2 \Delta\tilde{\ep}$~\cite{Jackson98}, and
hence the quantum analysis should apply.  Thus, by this informal
argument, we would expect the modal diameter to expand proportional to
$e^{C\lambda^2}$ for some constant $C$ (where $\lambda = 2\pi
c/\omega$ is the vacuum wavelength), but a more explicit proof would
be desirable.

\section{Concluding remarks}
\label{sec:conclusion}

We have demonstrated sufficient conditions for the existence of
cutoff-free guided modes for general microstructured dielectric
fibers, periodic in either or both the $z$ direction and in the
transverse plane. The results are a generalization of previous results
on the existence of such modes in fibers with a homogeneous cladding
index. Our theorem allows one to understand the guidance in many very
complicated structures analytically, and enables one to rigorously
guarantee guided modes in many structures (especially those where
$\dep \geq 0$ everywhere) by inspection.  There remain a number of
interesting questions for future study, however, some of which we
outline below.

Our \eqref{general-cond} is a \emph{sufficient} condition for
index-guided modes, but it is certainly not \emph{necessary} in
general: even when it is violated, one can have guided modes with a
cutoff (as for W-profile fibers~\cite{Kawakami74} or waveguides on
substrates~\cite{Hunsperger82, Chen06}), or other types of guided modes (such as
bandgap-guided
modes~\cite{Russell03,Bjarklev03,Zolla05,Joannopoulos08}).  However,
these other types of guides modes in dielectric waveguides have a
long-wavelength cutoff, so one can pose the question: is
\eqref{general-cond} a necessary condition for \emph{cutoff-free}
guided modes (where $\vec{D}_c$ is given by the long-wavelength limit
of the cladding fundamental mode) in dielectric waveguides (as opposed
to TEM modes in metallic coaxial waveguides, which also have no
cutoff~\cite{Kong75})?  Based on theoretical reasoning and some
numerical evidence, we suspect that the answer is \emph{no}, but that
it may be possible to modify \eqref{general-cond} to obtain a
necessary condition. In particular, the variational theorem is closely
related to first-order perturbation theory: if one has a small
perturbation $\dep$ and substitutes the unperturbed field into the
Rayleigh quotient, the result is the first-order perturbation in the
eigenvalue.  However, when $\dep$ is large, even if the volume of the
perturbation is small, perturbation theory requires a correction due
to the electric-field discontinuity at the
interface~\cite{Johnson05:bump}.  In the long-wavelength limit,
perturbation theory is corrected by computing the quasi-static
polarizability of the perturbation~\cite{Johnson05:bump}, and we
conjecture that a similar correction to our trial field may allow one
to derive a necessary condition for the absence of a cutoff.
\Eqref{general-cond} is still a sufficient condition (the variational
theorem still holds even with a suboptimal trial function), but the
preceding considerations predict that it will become farther from a
necessary condition for the absence of a cutoff as $\dep$ is
increased, and this prediction seems to be confirmed by preliminary
numerical experiments with W-profile fibers.

Let us also mention five other interesting directions to pursue.
First, \citeasnoun{Bamberger90} actually proved a somewhat stronger
condition than \eqref{general-cond} for homogeneous claddings, in that
they showed the existence of guided modes when the integral was $\leq
0$ (and $\dep >0$ in some region) rather than $< 0$ as in our
condition. Although the $=0$ case seems unlikely to be experimentally
or numerically significant, we suspect that a similar generalization
should be possible for our theorem (re-weighting the integrand to make
it negative and then taking a limit as in \citeasnoun{Bamberger90}).
Second, as discussed in \secref{substrates-etc}, it would be desirable
to develop a sufficient condition at a fixed $\omega$ rather than at a
fixed $\beta$, although we are not sure whether this is possible.
Third, one would like a more explicit confirmation of the argument, in
\secref{substrates-etc}, that the modal diameter should asymptotically
increase exponentially with the square of the wavelength.  Fourth, it
might be interesting to consider the case of ``Bragg fiber''
geometries consisting of ``periodic'' sequences of concentric
layers~\cite{Yeh78}, which are not strictly periodic because the layer
curvature decreases with radius. Finally, as we mentioned in
\secref{theorem}, it is possible to extend the theorem to a condition
for \emph{two} guided modes in many cases where the cladding
fundamental mode is doubly degenerate, and we are currently preparing
another manuscript describing this result along with conditions for
truly single-mode (``single-polarization'') waveguides.

\section*{Acknowledgements}

This work was supported in part by the US Army Research Office under
contract number W911NF-07-C-0002. The information does not necessarily
reflect the position or the policy of the Government and no official
endorsement should be inferred.  We are also grateful to
M.~Ghebrebrhan and G.~Staffilani at MIT for helpful discussions.

\section*{Appendix: All Rayleigh-quotient terms}

In this appendix, we provide an exhaustive listing of all the terms
that appear when the trial function [\eqref{general-trial}] is
substituted into \eqref{gen-var-condition} (the condition to be
satisfied, a rearrangement of the Rayleigh quotient bound).  Since the
terms that contain $\Delta$ [category~(i)] were already fully
analyzed in \secref{general} (since for these terms the limits could
be trivially interchanged), we consider only the remaining terms
involving $\ep_c(\vec{r})$.  More specifically, the only non-trivial term to
analyze is the $\Delta$-free part of the left-most integral in
\eqref{gen-var-condition}:
\begin{equation}
\begin{split}
&\int \vec{H}^*\cdot \left(\nabla_{\beta}\times \ep_c^{-1}\nabla_{\beta}\times\vec{H}\right) \\
&=\int \vec{H}^*\cdot \left(\nabla_{\beta}\times \ep_c^{-1}\gamma\nabla_{\beta}\times{\vec{H}_c}\right)+\int \vec{H}^*\cdot \left(\nabla_{\beta}\times \ep_c^{-1}\nabla\gamma\times{\vec{H}_c}\right)\\
&\quad{}+\int \vec{H}^*\cdot \left(\nabla_{\beta}\times \ep_c^{-1}\nabla_\beta\times\left(\nabla\gamma\times{\vec{A}_c}\right)\right) .
\end{split}
\end{equation}
We have already seen, in \eqref{eigen-cancellation}, that the first
term breaks down into a term that cancels $\frac{\omega_c^2}{c^2}\int
\vec{H}^*\cdot\vec{H}$ in \eqref{gen-var-condition}, via the
eigen-equation, and two other terms. Removing the terms cancelled by
the eigenequation, and substituting $-i\frac{\omega_c}{c}\vec{E}_c$
for $\ep_c^{-1} \nabla_\beta\times\vec{H}_c$ (Amp{\`e}re's law), we have:
\begin{equation}
\begin{split} 
&\quad{}-\frac{\omega_c^2}{c^2}\int \vec{H}^*\cdot\left(\nabla\gamma\times \vec{A}_c\right)+\int \vec{H}^*\cdot\left[\nabla\gamma\times \left(-i\frac{\omega_c}{c} \vec{E}_c\right)\right]\\
&\quad{}+\int \vec{H}^*\cdot \nabla_{\beta}\times\ep_c^{-1}\left[\nabla\gamma\times \vec{H}_c +\nabla_{\beta}\times\left(\nabla\gamma\times \vec{A}_c\right)\right]\\
&=-\frac{\omega_c^2}{c^2}\int\left[\gamma \vec{H}_c+\nabla\gamma\times \vec{A}_c\right]^*\cdot \left(\nabla\gamma\times \vec{A}_c\right)-i\frac{\omega_c}{c}\int \gamma \nabla\gamma\cdot\left(\vec{E}_c\times \vec{H}_c^*\right)\\
&\quad{}-i\frac{\omega_c}{c} \int \left(\nabla\gamma\times \vec{A}_c\right)^*\cdot\left(\nabla\gamma\times \vec{E}_c\right)+\int \left(\gamma \nabla_\beta\times \vec{H}_c\right)^*\cdot \ep_c^{-1}\left[\nabla\gamma\times \vec{H}_c +\nabla_{\beta}\times\left(\nabla\gamma\times \vec{A}_c\right)\right]\\
&\quad{}+\int \left(\nabla\gamma\times \vec{H}_c\right)^*\cdot \ep_c^{-1}\left[\nabla\gamma\times \vec{H}_c +\nabla_{\beta}\times\left(\nabla\gamma\times \vec{A}_c\right)\right]\\
&\quad{}+\int\left(\nabla_\beta\times\nabla\gamma\times \vec{A}_c\right)^*\cdot\ep_c^{-1}\left[\nabla\gamma\times \vec{H}_c +\nabla_{\beta}\times\left(\nabla\gamma\times \vec{A}_c\right)\right]\\
&=-\frac{\omega_c^2}{c^2}\int\gamma \vec{H}_c^*\cdot\left(\nabla\gamma\times \vec{A}_c\right)-\frac{\omega_c^2}{c^2}\int\left\|\nabla\gamma\times \vec{A}_c\right\|^2-i\frac{\omega_c}{c}\int \gamma \nabla\gamma\cdot\left(\vec{E}_c\times \vec{H}_c^*\right)\\
&\quad{}-i\frac{\omega_c}{c} \int \left(\nabla\gamma\times \vec{A}_c\right)^*\cdot\left(\nabla\gamma\times \vec{E}_c\right)+i\frac{\omega_c}{c}\int\gamma \vec{E}_c^*\cdot \left(\nabla\gamma\times \vec{H}_c\right)\\
&\quad{}+i\frac{\omega_c}{c}\int\gamma\left(\nabla_\beta\times \vec{E}_c\right)^*\cdot\left(\nabla\gamma\times \vec{A}_c\right)+i\frac{\omega_c}{c}\int\left(\nabla\gamma\times \vec{E}_c\right)^*\cdot\left(\nabla\gamma\times \vec{A}_c\right)\\
&\quad{}+\int \left(\nabla\gamma\times \vec{H}_c\right)^*\cdot\ep_c^{-1}\left(\nabla\gamma\times \vec{H}_c\right)\\
&\quad{}+\left(\int\left(\nabla\gamma\times \vec{H}_c\right)^*\cdot \ep_c^{-1}\left[\nabla_{\beta}\times\left(\nabla\gamma\times \vec{A}_c\right)\right]+\mathrm{c.c.}\right)\\
&\quad{}+\int\left(\nabla_\beta\times\left(\nabla\gamma\times \vec{A}_c\right)\right)^*\cdot\ep_c^{-1}\left(\nabla_\beta\times\left(\nabla\gamma\times \vec{A}_c\right)\right).
\end{split}
\end{equation}
Above, the first ``$=$'' step is obtained by substituting the trial
function for $\vec{H}$, integrating some of the $\nabla_\beta\times{}$
operators by parts, and distributing the derivatives of
$\gamma\vec{H}_c$ by the product rule.  The second step is obtained by
using Amp{\`e}re's law again, combined with integrations by parts and the
product rule; ``c.c.'' stands for the complex conjugate of the
preceding expression.  Continuing, we obtain:
\begin{equation}
\begin{split}
&=-\frac{\omega_c^2}{c^2}\int\gamma \vec{H}_c^*\cdot\left(\nabla\gamma\times \vec{A}_c\right)-\frac{\omega_c^2}{c^2}\int\left\|\nabla\gamma\times \vec{A}_c\right\|^2-2i\frac{\omega_c}{c}\int \gamma \nabla\gamma\cdot\Re\left\{\vec{E}_c\times \vec{H}_c^*\right\}\\
&\quad{}+\left(i\frac{\omega_c}{c} \int \left(\nabla\gamma\times \vec{A}_c\right)\cdot\left(\nabla\gamma\times \vec{E}_c\right)^*+\mathrm{c.c.}\right)+\frac{\omega_c^2}{c^2}\int\gamma \vec{H}_c^*\cdot\left(\nabla\gamma\times \vec{A}_c\right)\\
&\quad{}+\int \left(\nabla\gamma\times \vec{H}_c\right)^*\cdot\ep_c^{-1}\left(\nabla\gamma\times \vec{H}_c\right) +\left(\int\left(\nabla\gamma\times \vec{H}_c\right)^*\cdot \ep_c^{-1}\left[\nabla_{\beta}\times\left(\nabla\gamma\times \vec{A}_c\right)\right]+\mathrm{c.c.}\right)\\
&\quad{}+\int\left(\nabla_\beta\times\left(\nabla\gamma\times \vec{A}_c\right)\right)^*\cdot \ep_c^{-1}\left(\nabla_\beta\times\left(\nabla\gamma\times \vec{A}_c\right)\right)\\
\end{split}
\end{equation}
In obtaining this expression, we have grouped terms into
complex-conjugate pairs and used Faraday's law to replace
$\nabla_\beta\times\vec{E}_c$ with $i\frac{\omega_c}{c}\vec{H}_c$.  At
this point, we have two $\frac{\omega_c^2}{c^2}\int
\gamma\vec{H}_c^*\cdot\left(\nabla\gamma\times\vec{A}_c\right)$ terms
that exactly cancel.  All of the remaining terms, except for
$-i\frac{\omega_c}{c}\int \gamma
\nabla\gamma\cdot\Re\left\{\vec{E}_c\times \vec{H}_c^*\right\}$, are
multiples of two first or higher derivatives of $\gamma$,
corresponding to category~(iv) and~(v) terms, which we proved to vanish in
\secref{general}.

The only remaining term is the $\vec{E}_c\times\vec{H}_c^*$ term, in
category~(iii).  This term is identically zero (for any $\alpha\geq0$)
because it is purely imaginary, whereas all of the other terms are
purely real and the overall expression must be real.  More explicitly:
\begin{equation}
\begin{split}
&-2i\frac{\omega_c}{c}\int \gamma \nabla\gamma\cdot\Re\left\{\vec{E}_c\times \vec{H}_c^*\right\}\\
&\quad=-2i\frac{\omega_c}{2c}\int \nabla\gamma^2\cdot\Re\left\{\vec{E}_c\times \vec{H}_c^*\right\}\\
&\quad=-i\frac{\omega_c}{c}\int \nabla\cdot\left(\gamma^2\Re\left\{\vec{E}_c\times \vec{H}_c^*\right\}\right)+i\frac{\omega_c}{c}\int\gamma^2 \nabla\cdot\left(\Re\left\{\vec{E}_c\times \vec{H}_c^*\right\}\right)\\
\end{split}
\end{equation}
The first term of the last line is zero by the divergence theorem
(transforming it into a surface integral at infinity), since
$\gamma\to0$ at infinity.  For the second term, the integrand is the
divergence of the time-average Poynting vector
$\Re\left\{\vec{E}_c\times \vec{H}_c^*\right\}$, which equals the
time-average rate of change of the energy density~\cite{Jackson98}, which is
identically zero for any lossless eigenmode (such as the cladding
fundamental mode).

\end{document}